\documentclass[prl,aps,superscriptaddress]{revtex4}
\usepackage{times} 
\usepackage{amsfonts}
\usepackage{amssymb}
\usepackage[dvips]{graphicx}
\usepackage[T1]{fontenc}
\usepackage[latin2]{inputenc}

\usepackage{amsmath}
\usepackage{bbm}

\def\>{\rangle}
\def\<{\langle}

\begin{document}


\title{Linear dynamical entropy and free-independence for quantized linear maps
on the torus}

\author{Monika Pogorzelska and Robert Alicki}
\affiliation{Institute of Theoretical Physics and Astrophysics,
University of Gda\'nsk, \\ Wita Stwosza 57, PL 80-952 Gda\'nsk,
Poland\\dokmpo@univ.gda.pl, fizra@univ.gda.pl}

\begin{abstract}

We study the relations between the averaged linear entropy production in periodically measured quantum systems and ergodic properties of their classical counterparts. Quantized linear automorphisms of the torus, both classically chaotic and regular ones, are used as examples. Numerical calculations show different entropy production regimes depending on the relation
between the Kolmogorov-Sinai entropy and the measurement entropy. The hypothesis of free independence relations between the dynamics and measurement proposed to explain the initial constant and maximal entropy production is tested numerically for those models.

\end{abstract}

\maketitle

\section{Introduction}
In the last decade the relations between ergodic properties of classical systems and the dynamical phenomena for the corresponding quantized systems were discussed in a number of papers.
In particular the sensivity of the time evolution of quantum open systems with respect to external perturbations has been studied in terms of entropic quantities, purity of the state or fidelity in the spin-echo settings \cite{al,mon,emer,zin,all,cas,slom,patt,gar,bian,amp}. In most cases the external perturbation is modelled by the irreversible quantum dynamics combined with a unitary (Hamiltonian ) self-dynamics of the system.
As a result of  numerical analysis for different models there is a quite strong evidence that for classically chaotic systems with the Kolmogorov-Sinai entropy $h_{KS}$ the "entropy production" rate for their quantum counterparts is equal to $h_{KS}$  on the semiclassical time scale $\tau_{sem} \simeq \log N /h_{KS}$ and under certain technical conditions. Here $N$ is the dimension of the "effective" Hilbert space consisting of physically admissible states under certain given conditions.
\par
It is interesting that as the "entropy" we can put quite different quantities: 

1) von Neumann entropy of the initially pure state \cite{all},

2) Shannon entropy computed for diagonal elements in a fixed orthonormal basis \cite{cas},

3) Coherent states Shannon entropy \cite{slom},

4) Quantum dynamical entropy expressed in terms of multitime correlation matrices \cite{al},

6) The quantities of above with the Shannon (von Neumann) entropy replaced by the Renyi $\alpha$- entropies \cite{patt,gar,bian,amp}.
\par
It is also clear that the entropy production rate (EnPR) given by $h_{KS}$ can be observed only in a certain regime. Namely, for the strong coupling to an environment (measuring apparatus) the interaction with it should dominate over the system's self-dynamics, while
for the weak interaction the entropy production must be restricted by the interaction magnitude. However, even then we observe very often a constant EnPR over certain periods of time. Therefore, we still need more information about the onset and coexistence of the different regimes of entropy production. This knowledge is particularly important for the field of quantum control and quantum information processing. The efficient
quantum information processing needs time scale of the order of $\log$(dimension of Hilbert 
space) - exactly the time scales of the observed constant EnPR. On the other hand EnPR determines the averaged increase of error in the output state of the system per time unit.
\par
As suggested in \cite{all} the constant EnPR determined by the structure of the map describing
measurement process can be explained by a certain statistical independence, called {\emph{free independence}, between the unitary map governing the self-dynamics and the self-adjoint operator defining the measured system's observable. If such a hypothesis is true then one could consider a more general notion of \emph{quantum chaos} which is independent of any  classical
limit but depends only on the relations between dynamics and measurement.
\par
In the present work we address the question of the onset and coexistence of different regimes
of entropy production and the validity of  the {\emph{free independence hypothesis } in discrete-time quantum dynamical system periodically perturbed by projective von Neumann measurement. We choose quantized linear maps on the torus, both hiperbolic (chaotic) and elliptic (regular) ones, and the linear entropy in the ALF scheme \cite{all,alfan}. Such a combination of
systems and  parameters characterizing sensivity of the dynamics has not been used before and therefore enriches the existing "phenomenology".

\section{General scheme}
In the following we discuss a quantum system described by a finite dimensional Hilbert space $\mathbb{C}^{N}$ with discrete in time (stroboscopic) unitary self-dynamics
\begin{equation}
\rho \rightarrow{\cal U}^n\rho  = U^n \rho (U^{\dagger})^n \ ,\ U U^{\dagger}= U^{\dagger}U =\mathbbm{1}, \, n=0,1,2,....
\label{uni}
\end{equation}
If the system interacts with environment, then its evolution is in general irreversible, and described by a completely positive dynamical map. We consider a simplified model of environment represented by the von Neumann projective measurement which leads to the dynamical map 
\begin{equation}
\rho \rightarrow{\cal P}\rho  = \sum_{k=1}^K P_k \rho P_k \ ,\ P_k= P_k^2 = P_k^{\dagger}\ ,\ \sum_{k=1}^K P_k =\mathbbm{1} .
\label{mea}
\end{equation}
The evolution of the periodically measured system can be defined as follows
\begin{equation}
\rho(n) = \sum_{j_{n}}^K...\sum_{j_{1}}^K P_{j_{n}}U...P_{j_{1}}U\rho(0) U^\dagger P_{j_{1}}...U^\dagger P_{j_{n}},
\label{evol}
\end{equation}
or in a short-hand notation as
\begin{equation}
\rho(n)=[ {\cal P} {\cal U}]^{n} \rho(0).
\label{rhon}
\end{equation}
Now the speed of decoherence measured either by entropic quantities, purity or fidelity
of the final (output) state can be a natural measure of stability of the evolution.
However, the output state depends on the input state of the system $\rho(0)$. To have the parameter independent of the initial state, we follow the construction used in the definition of the ALF quantum dynamical entropy \cite{all,alfan}. We introduce an ancillary system and assume that the system plus ancilla is in a maximally entangled state
\begin{equation}
|\Psi \>=\frac{1}{\sqrt{N}}\sum_{l=1}^{N}|l\>\otimes |l\> ,
\label{me}
\end{equation}
where $\{|l\>\}_{l=1}^{N}$ is a basis in $\mathbb{C}^N$, and the Hilbert space of the composed system is 
$\mathcal{H}=\mathbb{C}^N\otimes\mathbb{C}^N$.\\ 
Then we compute the final state of the composed system after $n$ evolution steps given by
the $N^2\times N^2$ density matrix
\begin{equation}
\Omega[n]=[ {\cal P} {\cal U}\otimes \mathbb{I}_{anc}]^{n}|\Psi \>\<\Psi |
\label{omegan}
\end{equation}
and consider production of the linear entropy
\begin{equation}
I[n] = -\mathrm{ln}\mathrm{Tr}\Omega^2[n] .   
\label{ren1}
\end{equation}
It is not difficult to show that $I[n]$ does not depend on the choice of the basis
$\{|l\>\}$ in (\ref{me}) and satisfies the bound \cite{all}
\begin{equation}
I[n]\leq \mathrm{min}\{ n\mathrm{ln}K, 2\mathrm{ln}N\}.
\label{bound}
\end{equation}
Quite often one observes
a constant and maximal EnPR equal to $\mathrm{ln}K$. Using the fact that the nonzero eigenvalues (including their degeneracies) of the $N^2\times N^2$ density matrix $\Omega[n]$ (\ref{omegan}) concide with the non-zero eigenvalues of the $K^n\times K^n$ correlation matrix \cite{all,alfan}
\begin{equation}
D[{i_1,i_2,...,i_n;j_1,j_2,...,j_n}]= \frac{1}{N}\mathrm{Tr}\bigl(P_{j_1}U^{\dagger}P_{j_2}U^{\dagger}\cdots P_{j_n}P_{i_n}\cdots UP_{j_2}UP_{j_1}\bigr)    
\label{corm}
\end{equation}
we conclude that the maximal EnPR implies that
\begin{equation}
D[{i_1,i_2,...,i_n;j_1,j_2,...,j_n}]= \frac{1}{N}\mathrm{Tr}\bigl(P_{j_1}U^{\dagger}P_{j_2}U^{\dagger}\cdots P_{j_n}P_{i_n}\cdots UP_{j_2}UP_{j_1}\bigr)= \frac{1}{K^n}\delta_{i_1 j_1} \delta_{i_2 j_2}\cdots\delta_{i_n j_n} . 
\label{cormax}
\end{equation}

\section{Free Independent Variables}
The definition of Free Independent Variables (FIV), introduced by Voiculescu in order to explain some problems 
occurring in von Neumann algebras, found widespread aplications in the theory of random matrices \cite{voi1,voi2}. Bridging those two theories
is the theorem according to which random matrices in the infinite dimension limit are FIV.
The notion of FIV itself is a generalization of statistical independence of random variables to the case when the variables do not commute.\\ 
Recall, in the classical probability theory we consider an (commutative) algebra of random variables and the probability measure which defines the average of any random variable  $f\mapsto \<f\>$. The random variables $f_1,f_2,...,f_n$ are called (statistically) independent if for any choice of polynomials  $w_1, w_2,...,w_n$ we have 
\begin{equation}
\<w_1(f_1)w_2(f_2)\cdots w_n(f_n)\> =\<w_1(f_1)\>\<w_2(f_2)\>\cdots \<w_n(f_n)\> .
\label{ind}
\end{equation}
In quantum probability the noncommutative operator $^{\star}$-algebra $\mathcal{A}$ contains quantum random variables (quantum observables) and a linear, positive and normalized functional $\phi$
(quantum state) defines a mean value of any quantum observable $A \mapsto \phi(A)$.
As quantum random variables in general do not commute, we can have a much more richer structure of non-equivalent notions generalizing classical statistical independence.
One of them is a \emph{free independence} which is defined as follows.\\
The collection of noncommutative  random variables $A_1, A_2,...,A_n$ is called free independent
if
\begin{equation}
\phi(w_1(A_{p_1})w_2(A_{p_2})\cdots w_m(A_{p_n})) = 0,\,\, m=0,1,2,...
\label{find}
\end{equation}
whenever $\phi(w_j(A_{p_j}))= 0$ and $p(j)\neq p(j+1)$ for all $j= 1,2,...,m$\, and \, $p(j)\in \{1,2,...,n\}$.\\

In the ref. \cite{all} it was argued that the hypothesis of free independence can be used to explain the phenomenon of maximal EnPR as given by (\ref{corm}) , (\ref{cormax}). Namely, taking $N\times N$ matrices as an algebra of quantum observables  with the state $\phi$ given by a normalised trace as in (\ref{corm}) we can consider the free-independence condition (\ref{find}) for a pair $\{U, A\}$ where $A = \sum a_j P_j$. Adding some simplifying conditions
\begin{equation}
\frac{1}{N}\mathrm{Tr} P_j = \frac{1}{K},\,\, j=1,2,...,K,\,\, \frac{1}{N}\mathrm{Tr} U^n = 0,\,\, n=1,2,...
\label{find1}
\end{equation}
one can show that the free independence (\ref{find}) of $\{U, A\}$ is equivalent to the following condition
\begin{equation}
C[n]\equiv C[r_1,r_2,...,r_n; k_1,k_2,...,k_n]=\frac{1}{N}\mathrm{Tr}\bigl(U^{r_1}Q_{k_1}U^{r_2}Q_{k_2}\cdots U^{r_n}Q_{k_n}) =0, \,\, n=1,2,...,\,\,
r_j = \pm 1,\pm 2,...,
\label{find2}
\end{equation}
with $ Q_k = P_k - \frac{1}{K}$, $ k=1,2,...,K$.
Indeed, using (\ref{find2}) one can compute the correlations (\ref{corm}) to obtain (\ref{cormax}). Obviously, as the maximal EnPR can be observed, at most, at the time scale $\tau_{max} = 2\mathrm{ln}N/\mathrm{ln}K$, the free independence condition (\ref{find2}) can be valid for $n \leq 2\mathrm{ln}N/\mathrm{ln}K$ only. Therefore it should be treated as an asymptotic condition for $N\to\infty$.

\section{Quantised linear maps on a torus}
The classical linear automorphism  of the torus, maps a two-dimensional torus into itself
\begin{equation}
\mathrm{x}^{'}=\mathrm{T}\mathrm{x} \,|\mathrm{mod} 1 ,\,\,\, \mathrm{x}=[x_1,x_2]\in (0,1]\times (0,1],
\end{equation}
here $t_{ik}\in \mathbb{Z}$ and we assume that $\mathrm {Det T}=1$.\\
If the condition $|\mathrm {Tr T}|>2$ holds then the map is called hiperbolic (Arnold cat map).
This mapping is based on stretching the unity square along the unstable direction (indicated by a greater eigenvalue 
$\lambda_{+}$ of the matrix T) and compressing it along the stable direction (corresponding to smaller eigenvalue $\lambda_{-}$). Kolmogorov-Sinai entropy of such a mapping is equal to
\begin{equation}
h_{KS}=\mathrm{ln} \lambda_{+}.
\end{equation}
In the following we choose the  matrix T
\begin{displaymath}
\mathrm{T}=\left[ \begin{array}{cc}
2&7\\
1&4\\
\end{array} \right]
\end{displaymath}
for which $h_{KS}=1,76$.\\

A quantized vesion  of the cat map \cite{han} on the $N$- dimensional Hilbert space
is given in a fixed basis by the $N\times N$ unitary matrix
\begin{equation}
U^{cat}_{rk}=\sqrt{\frac{1}{N}}\mathrm{exp}\{ \frac{-2\pi i}{N}(r^2+2k^2-kr) \}, \,\, r,k = 1,2,...N .
\label{hip}
\end{equation}
The elliptic version of the linear automorphism of the torus satisfies the condition $|\mathrm {Tr T}|<2$ which gives K-S entropy equal to zero. We use the particular matrix 
\begin{displaymath}
\mathrm{T}=\left[ \begin{array}{cc}
0&1\\
-1&0\\
\end{array} \right]
\end{displaymath}
with the quantum counterpart given by \cite{han}
\begin{equation}
U^{ell}_{rk}=\sqrt{\frac{i}{N}}\mathrm{exp}\{\frac{2\pi i}{N}rk \}, \,\, r,k = 1,2,...N. 
\label{ell}
\end{equation}
For comparison a unitary shift
\begin{equation}
U^{shift}|r>=|r+1>,\,\, |r+N>\equiv|r> 
\label{shift}
\end{equation}
is used also.\\
The von Neumann projective  measurement is defined in terms of the same basis as all unitary dynamical maps (\ref{hip}), (\ref{ell}) and (\ref{shift})
\begin{equation}
\mathbf{P}=\{P_{j}: P_{j}=\sum_{m\in (N_{j-1},N_{j}>}|m><m|,\,\, j = 1,2,...,K,\,\, 0=N_0<N_2 <\dots<N_K=N \}.
\label{dim1}
\end{equation}
We apply both, projector with equal dimensions, i.e. $N_j = (Nj/K)$
what is equivalent to partitioning the phase space of the system into equal parts, or with varying dimensions.\\

\section{ Results}
\subsection{Entropy production}

\begin{figure}[!h]
\begin{small}
\begin{tabular}{@{}l@{}l@{}}
(a) \includegraphics[height=4cm,width=3cm,angle=270]{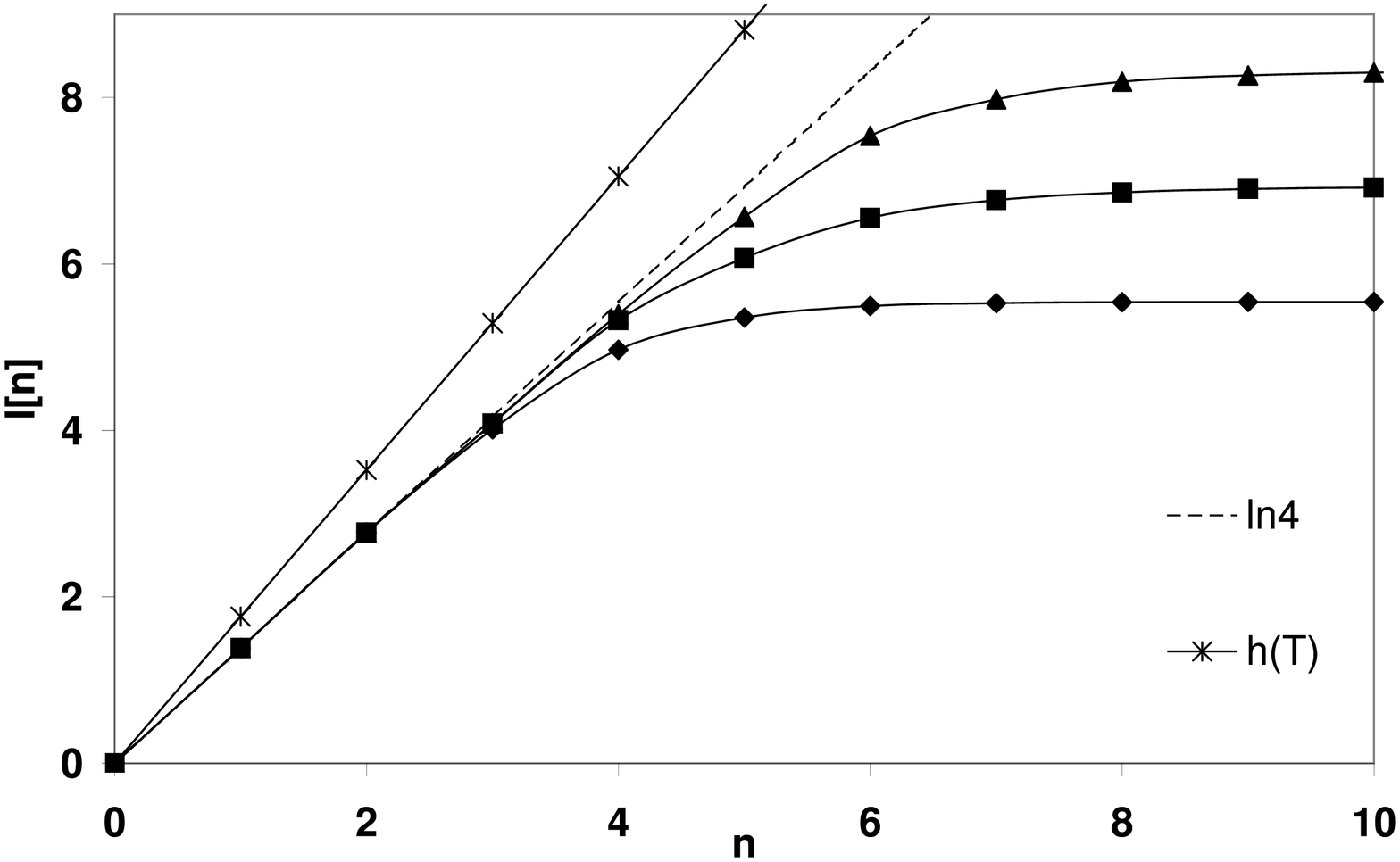}  \,\,\,\,\,\,\,\,\,\,\,\,  &(b)\includegraphics[height=4cm,width=3cm,angle=270]{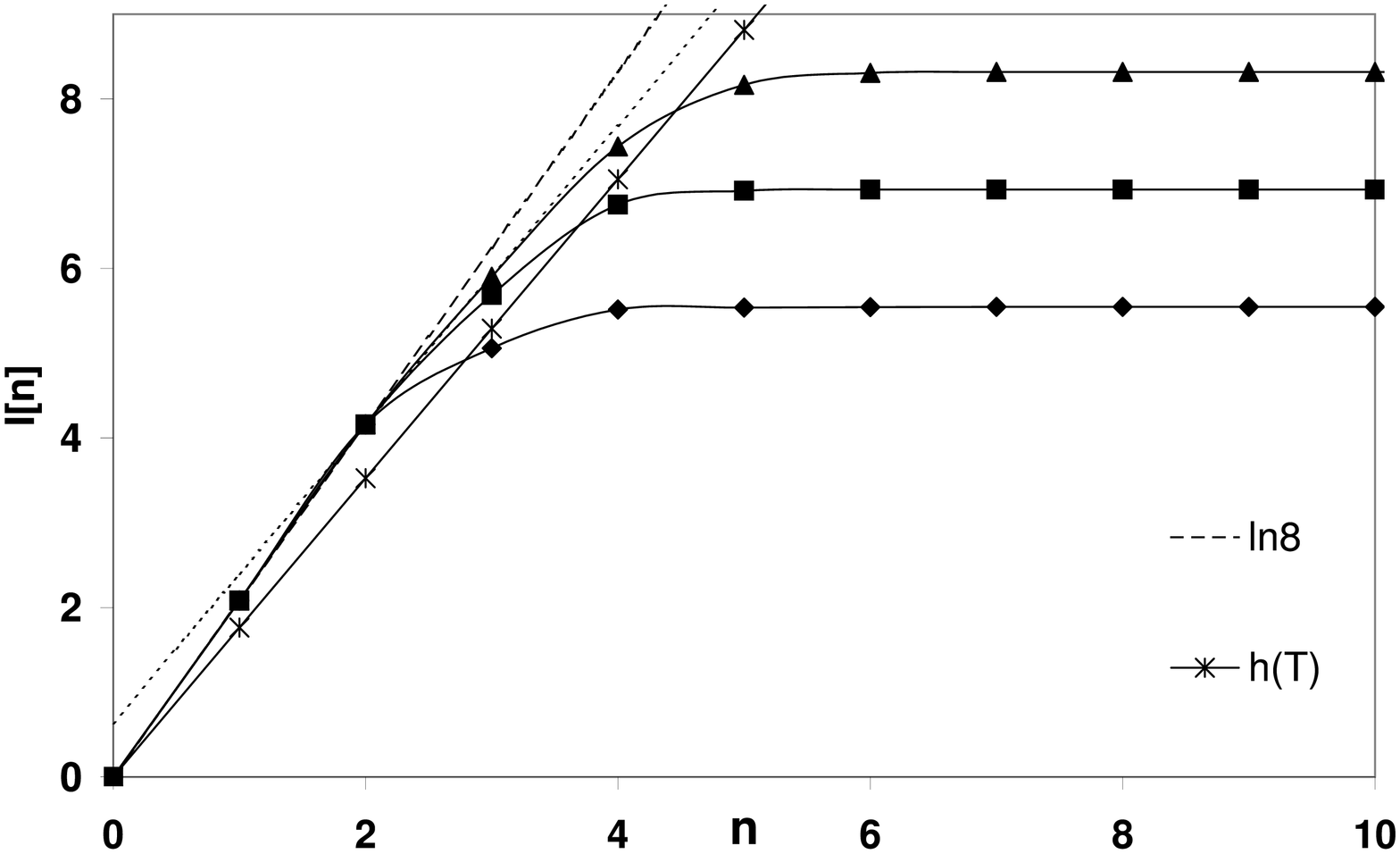}\\ [4cm]
(c) \includegraphics[height=4cm,width=3cm,angle=270]{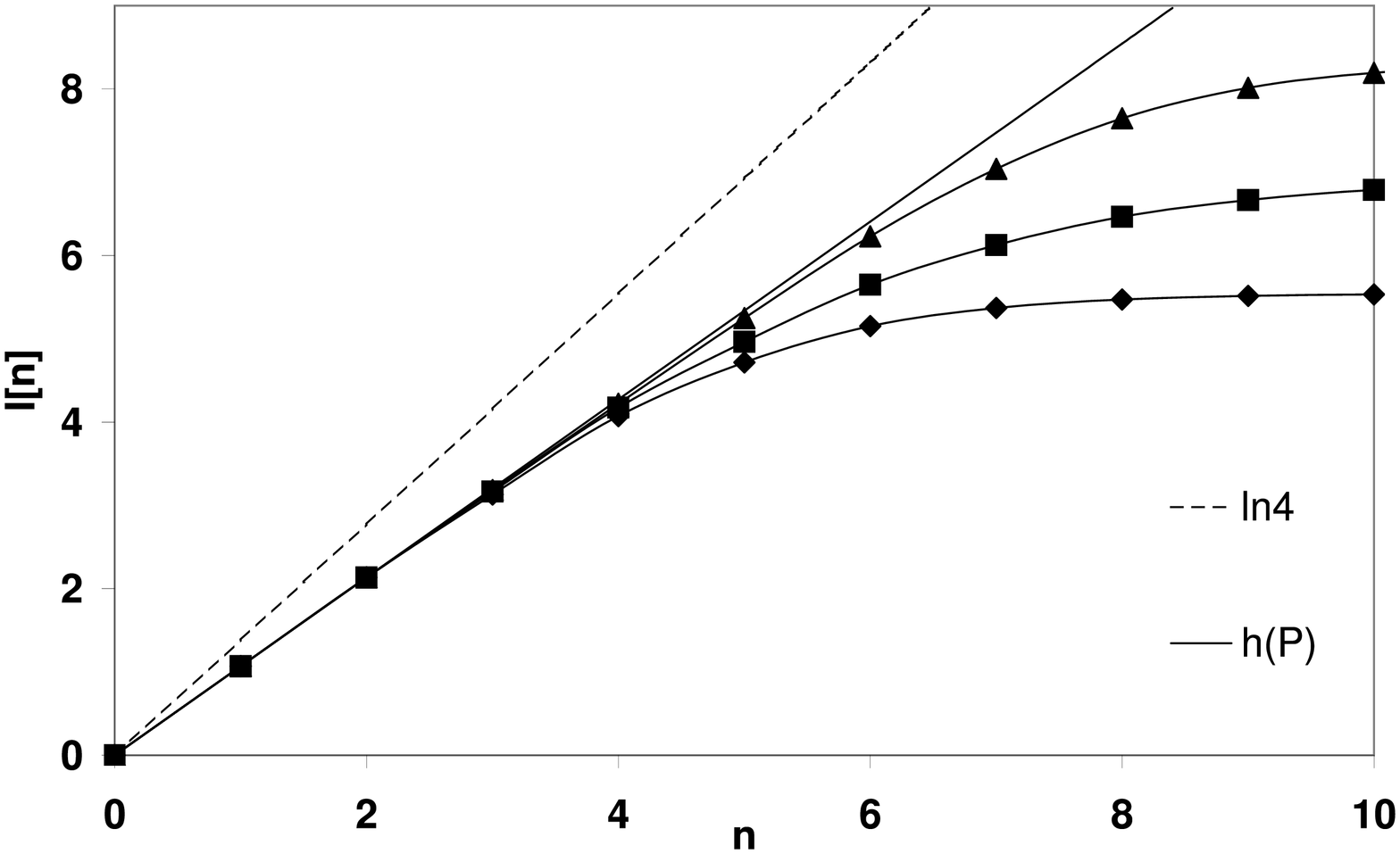}  \,\,\,\,\,\,\,\,\,\,\,\,   &(d)\includegraphics[height=4cm,width=3cm,angle=270]{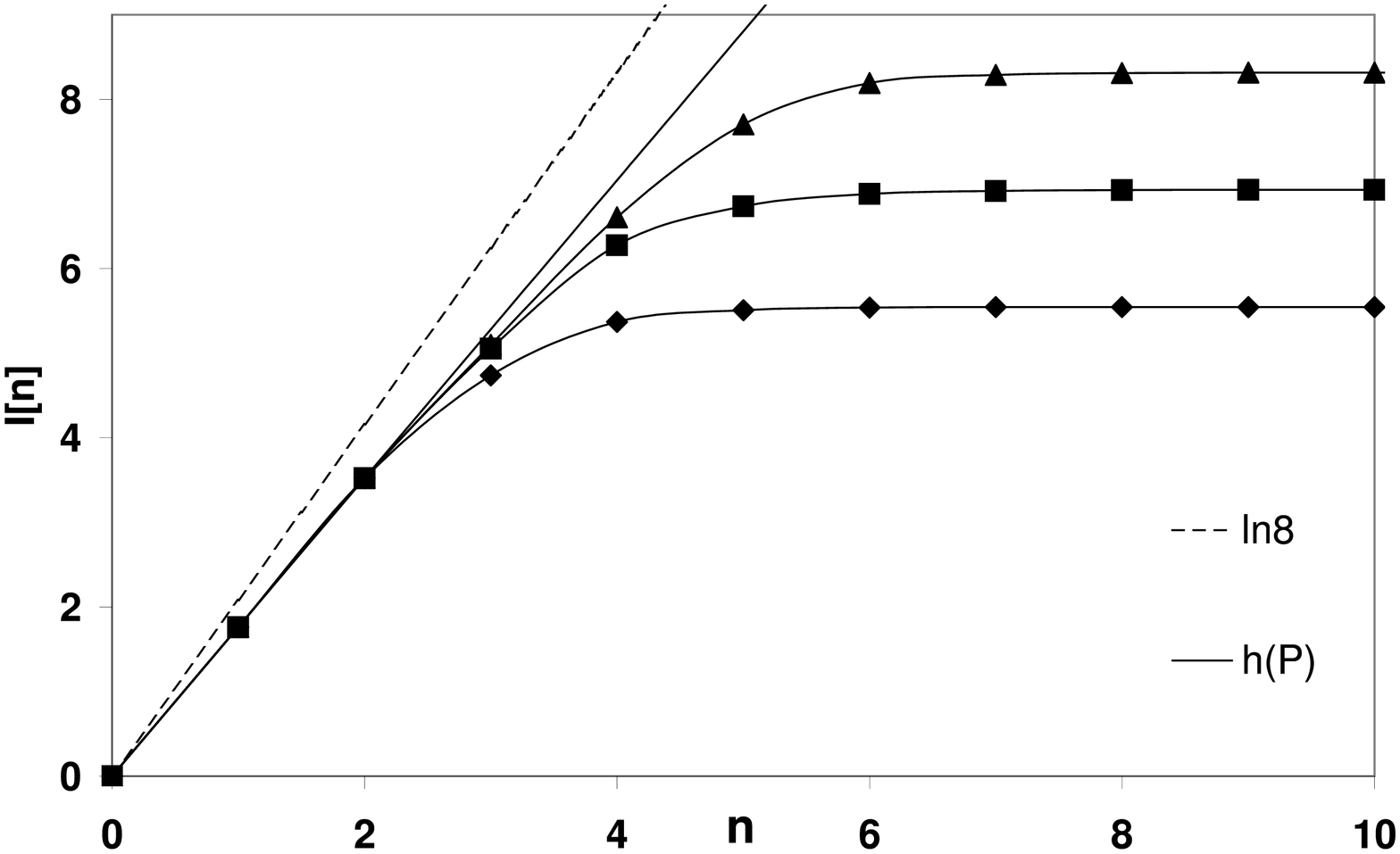}\\ [4cm]
\end{tabular}
\caption{\small  The graphs present the change of entropy in consequent stages of evolution, which was caused by 
measurements (a), (b) with projectors of equal dimensions; (c), (d) with projectors of varying dimensions, for different Hilbert space 
dimensions $N$=16 ($\blacklozenge$), $N$=32 ($\blacksquare$), $N$=64 ($\blacktriangle$).}
\end{small}
\end{figure}
We performed the numerical calculations for $N$=16, 32, 64 and $K$=4, 8. The results for the linear entropy I[n] are 
shown on the fig.1 and fig.2. Fig.1 shows the differences between measurements described with projections of equal and 
varying dimensions. If the phase space of the system is partitioned into equal parts and the partitioning is
coarse-grained enough ($\ln K> h_{KS}$), two regimes coexist. First, production of entropy is maximal i.e. equal to $\ln K$. Later, entropy production is given by the classical Kolmogorov-Sinai entropy $h_{KS}$ (see the dotted
line with the slope given by $h_{KS}$). For partitions with $\ln K \leq h_{KS}$ entropy production rate is given by $\ln K$ till the saturation at the maximal value $2\ln N$. For a measurement performed  with
projectors of varying dimensions, the initial entropy production is well described by the linear entropy which can be called \emph{measurement entropy}
\begin{equation}
h({\cal P})=-\mathrm{ln} \sum_{j=1}^{K} \frac{(\mathrm{Tr}P_j)^2}{N^2} ,
\label{lin}
\end{equation}
which better reflects  nonhomogeneity of the measurement than $\ln K $.\\ 
Fig. 2 presents a comparison of chaotic dynamics ($h_{KS}$=1,76) with two kinds of regular dynamics for $K$=4 and $K$=8. With the 
diamonds we have marked the increase of entropy in the system, the unitary evolution of which is described by the elliptic map (\ref{ell})
the triangles correspond to the shift (\ref{shift}).
Despite both systems being characterized in the classical limit by K-S entropy equal to zero, the entropy production in those systems behaves completely differently. For chaotic map we  observe a linear growth at the rate $\ln K  < h_{KS}$ followed by the rapid saturation at the maximal value $2\ln N$. For elliptic map, which is classically ergodic, after short linear growth the entropy approaches its maximal value at a much slower pace. For the non ergodic shift entropy reaches the saturation level equal to $\mathrm{ln}N$. 
\begin{figure}[!h]
\begin{small}
\begin{tabular}{@{}l@{}l@{}}
(a) \includegraphics[height=4cm,width=3cm,angle=270]{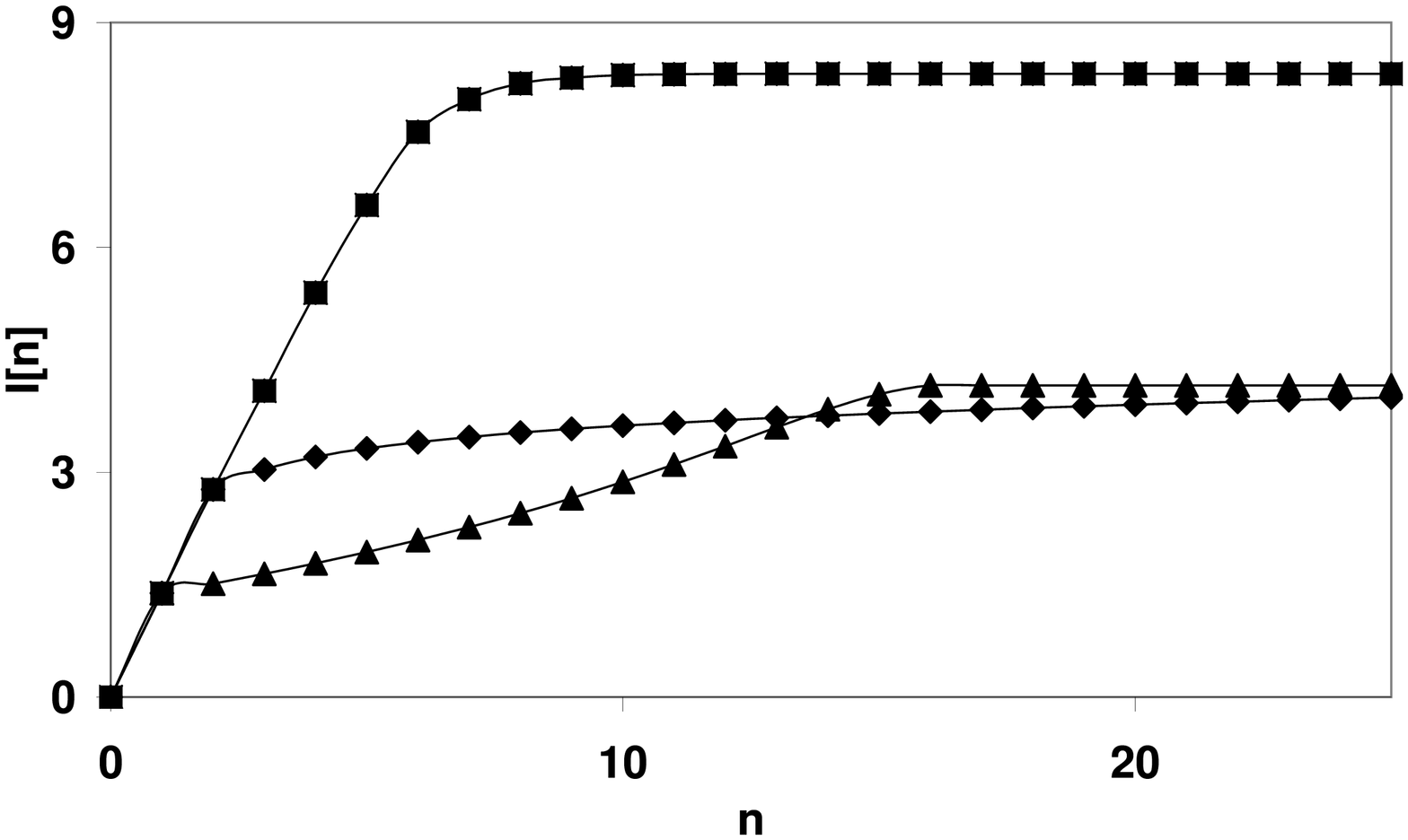}  &  \,\,\,\,\,\,\,\,\,\,\,\,   (b)\includegraphics[height=4cm,width=3cm,angle=270]{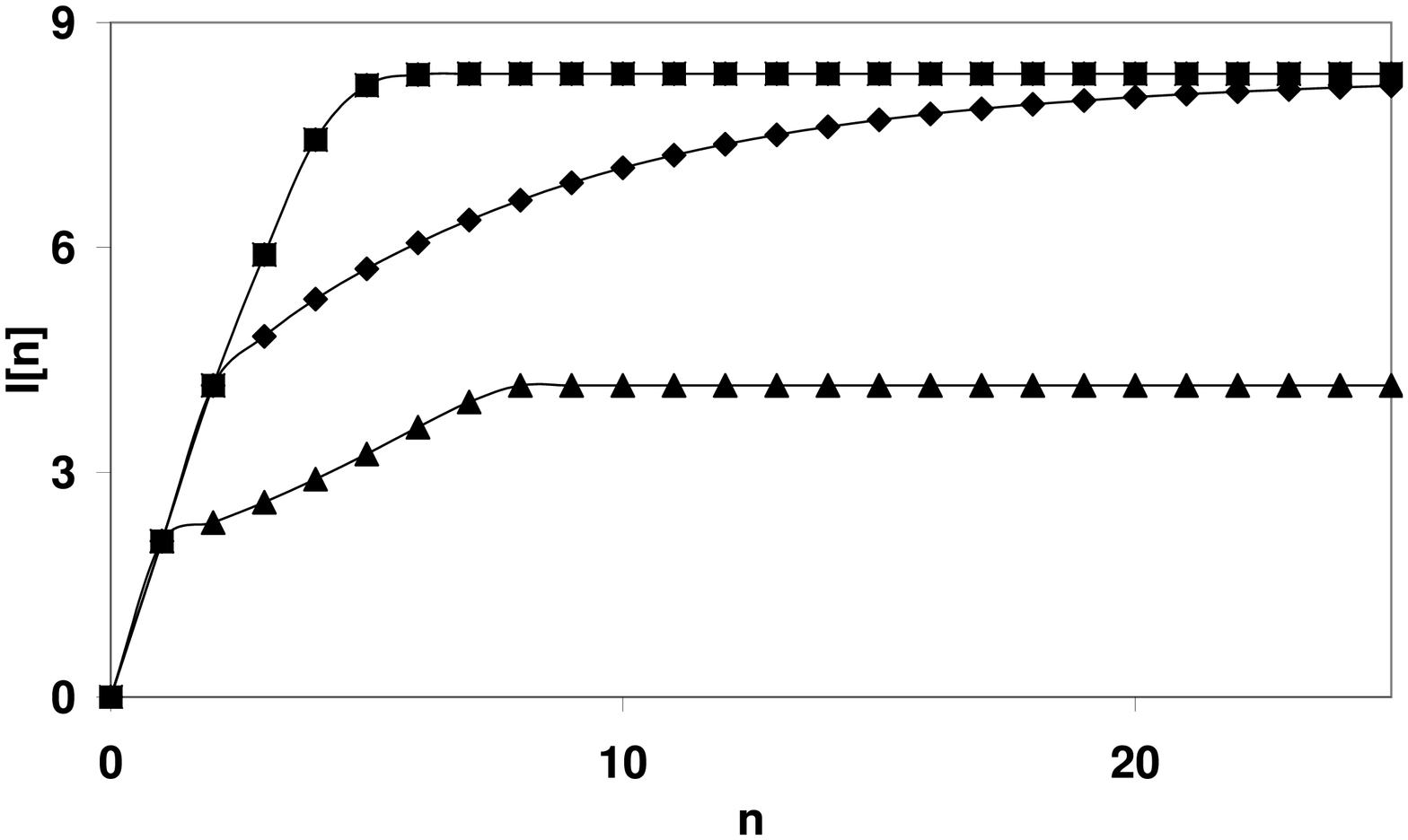}\\ [4cm]
\end{tabular}
\caption{\small The graphs present a comparison of the linear entropy as a function of n for quantum cat map ($\blacksquare$) ($h_{KS}$=1,76), with regular, elliptic mapping ($\blacklozenge$) 
and a shift ($\blacktriangle$) , for $N$=64 (a) $K$=4 (ln4<1,76), (b) $K$=8 (ln8>1,76).}
\end{small}
\end{figure}

\subsection{Free independence of U, $\mathbf{P}$}
\begin{figure}[!h]
\begin{small}
\begin{tabular}{@{}l@{}l@{}}
(a) \includegraphics[height=4cm,width=3cm,angle=270]{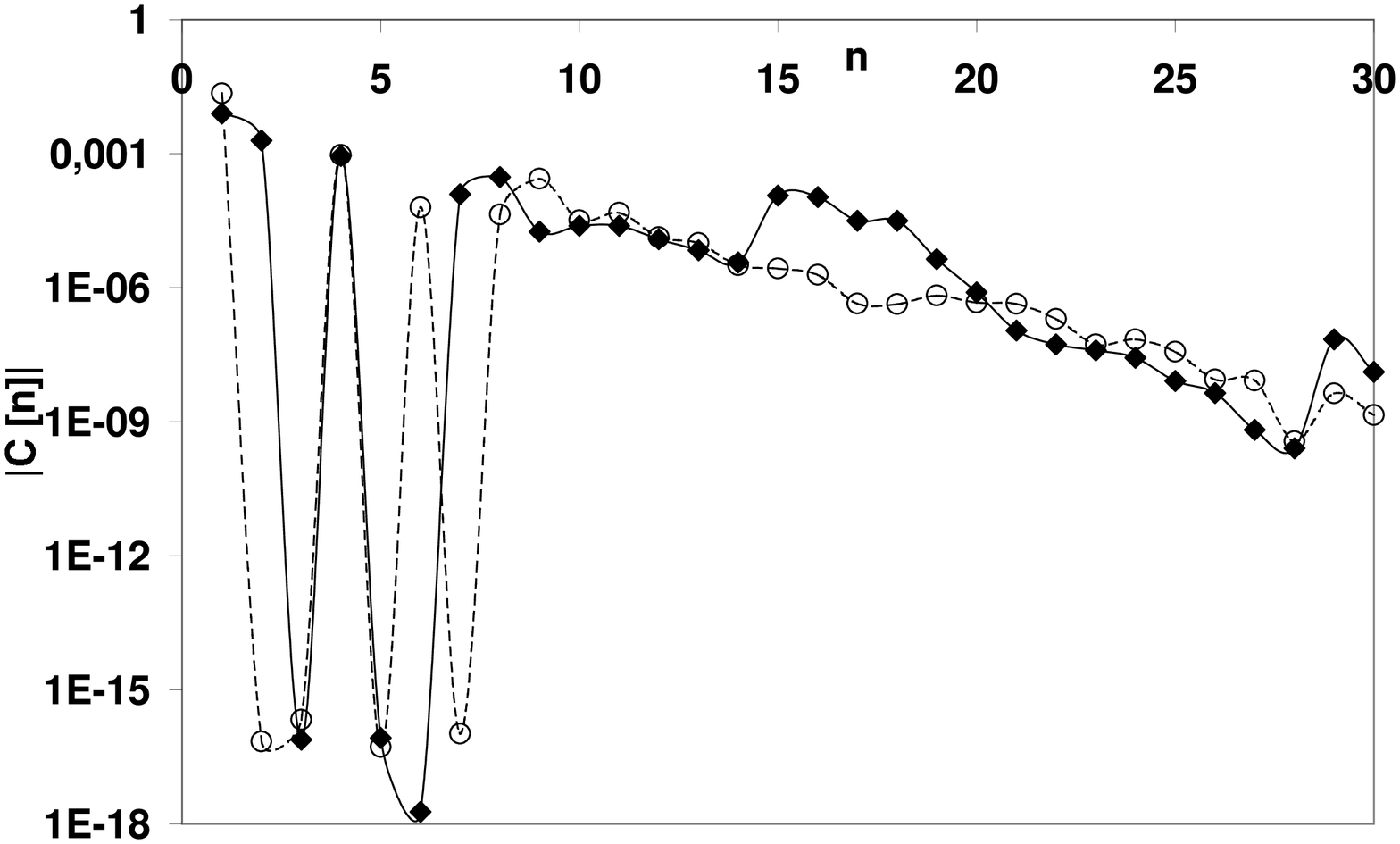}  &  \,\,\,\,\,\,\,\,\,\,\,\,   (b)\includegraphics[height=4cm,width=3cm,angle=270]{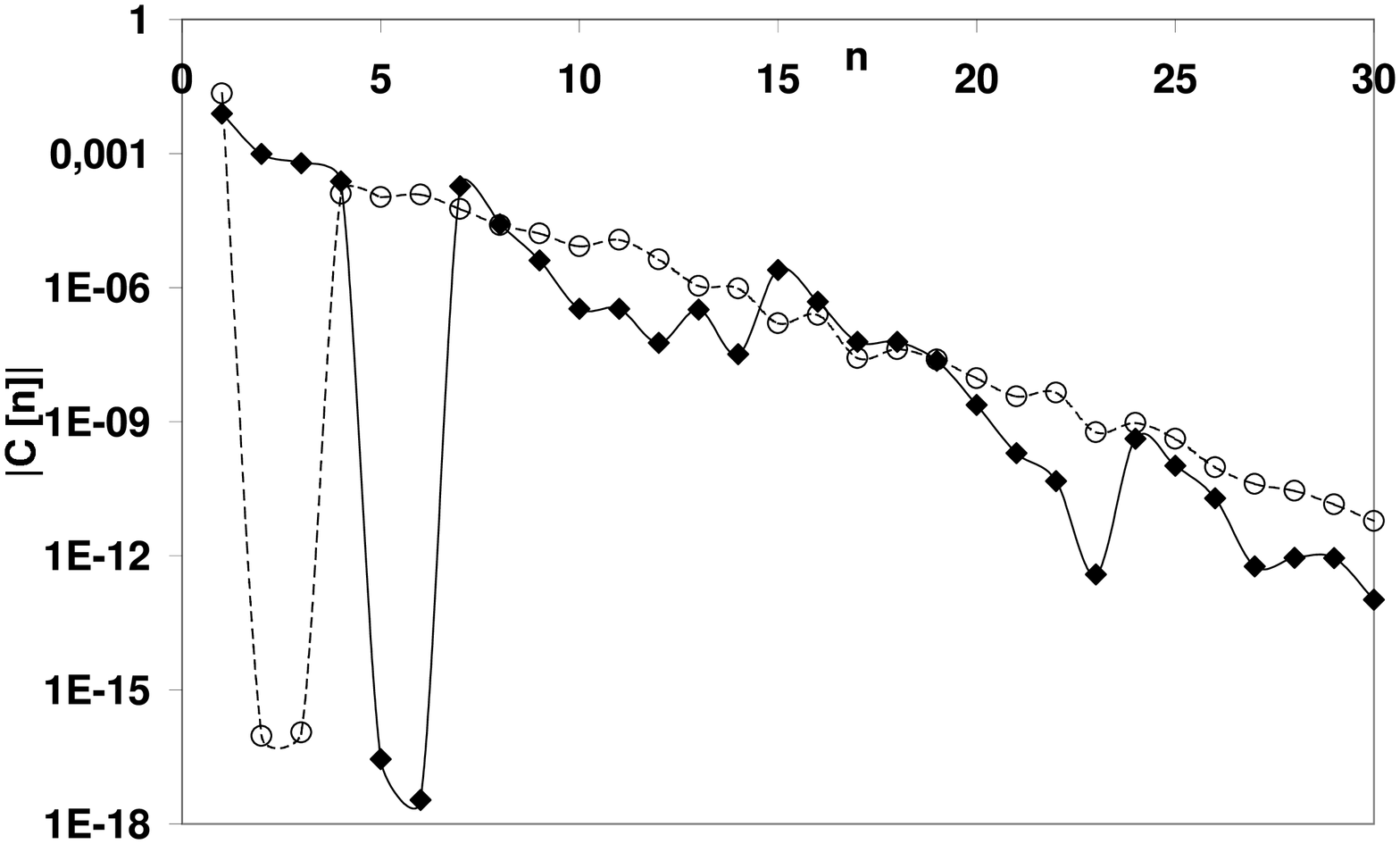}\\ [4cm]
\end{tabular}
\caption{\small The graphs show the absolute values of the correlation functions as a function of the "length" n, for quantum cat map ($\circ$) ($h_{KS}$=1,76) and  elliptic map ($\blacklozenge$) 
, for $N$=256 (a) $K$=4 (ln4<1,76), (b) $K$=8 (ln8>1,76) in logarithmic scale.}
\end{small}
\end{figure}

Fig. 3 shows the examples of computation of the absolute values of the correlation functions $|C[n]|$ (\ref{find2}) for the dynamical maps with $h_{KS}=1,76$ (cat map) and $h_{KS}=0$ (elliptic). It is important that for different random choices of sequences $\{m_1,j_1,...,m_n\}$ the qualitative picture is always the same. Firstly (fig. 4) we compare these function for a given $U$ and a given (properly scaled with $N$) partition ${\cal P}$ but with different Hilbert space dimensions $N$. For $n > 2\ln N/h({\cal P})$
the absolute values of $|C[n]|$ decay exponentially with the rate independent of $N$. On the other hand, for $n < 2\ln N/h({\cal P})$ functions $|C[n]|$ strongly fluctuate, but their  values are always fairly below the values extrapolated from the long-time exponential decay. This behavior is consistent with the free independent hypothesis in the region $n < 2\ln N/h({\cal P})$.\\
Fig. 5 exhibits the linear dependence of the long-time decay rates of $|C[n]|$ on the measurement entropy $h({\cal P})$. For $n > 2\ln N/h({\cal P})$ the simple Anzatz
\begin{equation}
|C[n]|\simeq \exp \{-\frac{1}{2}n A h({\cal P})\}
\label{anz}
\end{equation}
fits quite well with the values of $A$ close to 1. The exponential decay can be explained assuming that the quantum variables $F(m,j)=U^m P_j$, $m\neq 0$ behave like independent random variables in the standard sense (\ref{ind}) with the mean value 
\begin{equation}
|\<F(m,j)\>|\simeq \exp\bigl(-\frac{h({\cal P})}{2}\bigr) .
\label{anz1}
\end{equation}
Taking partition in $K$ equal parts we obtain from (\ref{anz1}) $ |\<F(m,j)\>|\simeq 1/\sqrt{K}$
what is in a good agreement with the estimation
\begin{equation}
\frac{1}{N}\mathrm{Tr}(F^{\dagger}(m,j)F(m,j))= \frac{1}{K}(1- 1/K)\simeq \frac{1}{K}  .
\label{anz2}
\end{equation}

\begin{figure}[!h]
\begin{small}
\begin{tabular}{p{8cm} p{1cm} p{8cm}}
\includegraphics[height=4cm,width=3cm,angle=270]{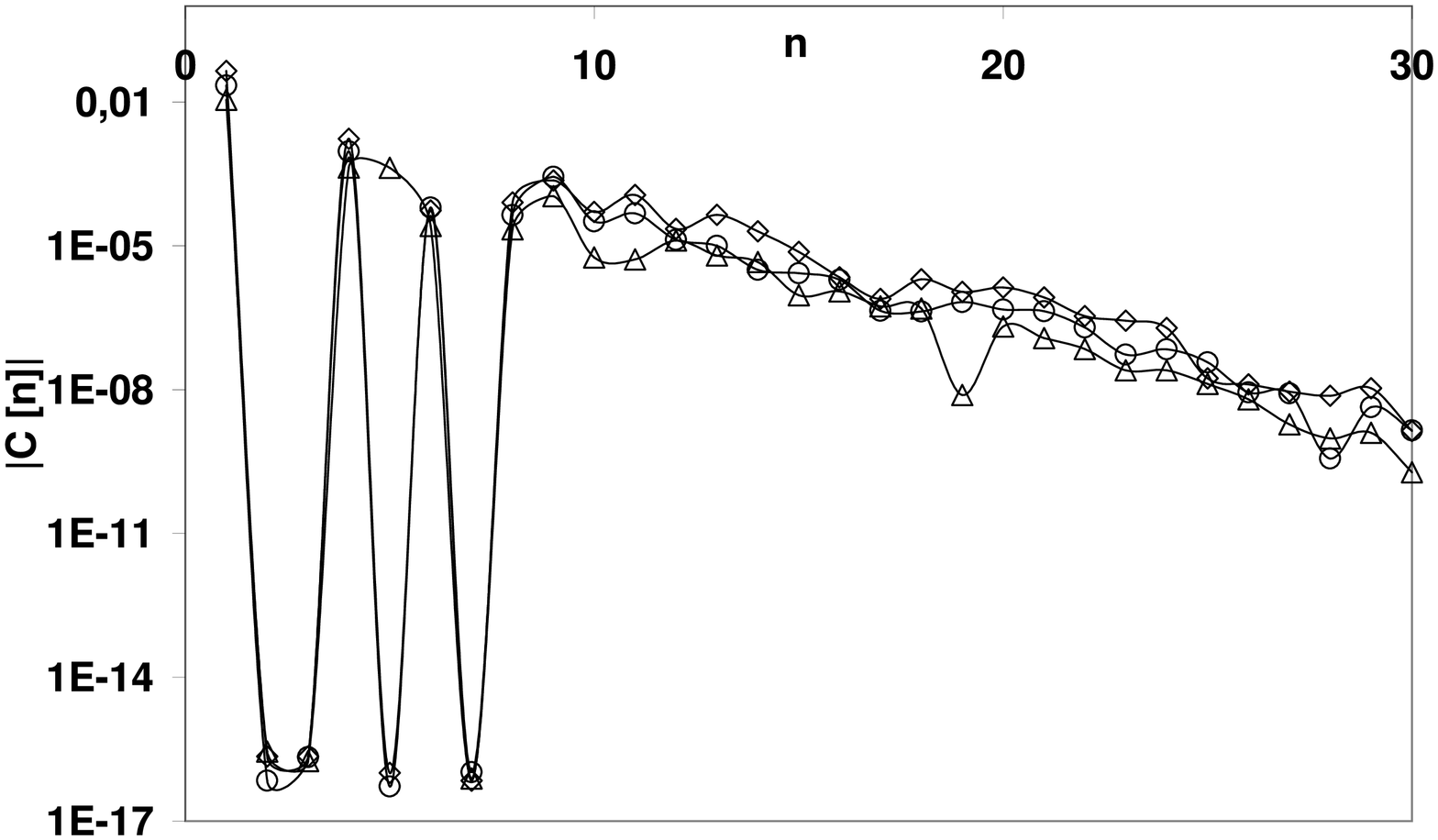}  &  &\includegraphics[height=4cm,width=3cm,angle=270]{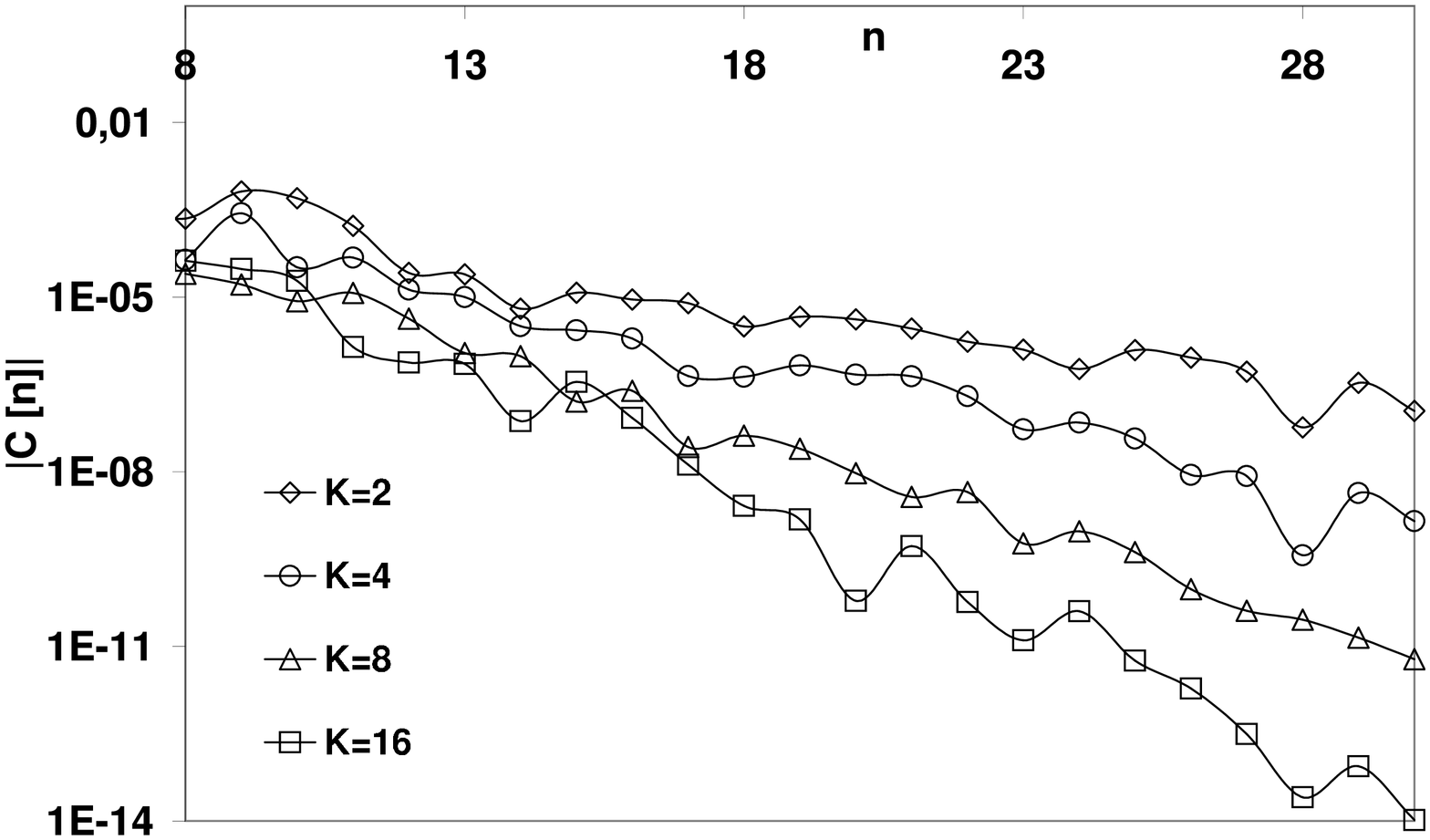} \\ [4cm]
\caption{\small The graph presents a comparison of the absolute values of the correlation function for different Hilbert spaces ($\diamond$) $N$=128,
($\circ$) $N$=256 and ($\bigtriangleup$)  $N$=512 for Arnolds cat map and $K$=4.} & &\caption{\small The absolute values of the correlation functions for diffrent $K$ and $n > 2\ln N/h({\cal P})$, $N$=256.}\\
\end{tabular} 
\end{small}
\end{figure}

\section{Conclusions}

The presented results together with the previous ones \cite{al,all,ben} can be used to draw a rather general picture of generic interrelations between decoherence rates in quantum systems perturbed by an environment and the ergodic properties of their classical counterparts. Assume that the classical system is chaotic with
the K-S entropy $h_{KS}$. Then its quantum counterpart interacting with the environment can be characterized on the quasiclassical time scale  $2\ln N /h_{KS}$ by a "steady state regime" with a constant EnPR equal to $h_{KS}$. How quickly or whether at all this steady state is reached depends on the magnitude and form of the interaction with an environment. In our model this magnitude is parametrized by the entropy of measurement (\ref{lin}). If it is larger than $h_{KS}$ then it determines the initial EnPR possibly followed by a transition to the steady state regime. If the measurement entropy is smaller then $h_{KS}$ we observe a maximal EnPR allowed by the measurement up to the saturation at $2\ln N$. In the case of ergodic classical systems with $h_{KS}=0$ the steady state regime with zero entropy production is approached asymptotically after initial stage of EnPR given by the measurement entropy. The more difficult problem is dependence on the "form" of the interaction. The results of \cite{all} shows that if the projections $P_j$ are chosen randomly the stationary regime is never reached, EnPR coincides with the measurement entropy up to the saturation at $2\ln N$.
\par
Therefore the following picture can be advocated. For a given quantum system with chaotic
classical counterpart there exist in the space of measurements (or couplings to an environment) 
a set of "fixed points" - the measurements with $h({\cal P})= h_{KS}$ and possessing semiclassical meaning. They correspond to \emph{generating partitions} in classical ergodic theory \cite{corn}. Around them there is a "basin of attraction" with measurements for which the dynamics approaches the steady state of EnPR $h_{KS}$. The measurements with $h({\cal P})$ much different from $h_{KS}$ or
randomly choosen with respect to a natural Hilbert space basis (such a "natural basis" should have a
classical interpretation in terms of simple phase-space functions) are beyond this basin. 
For them we expect EnPR given by  $h({\cal P})$ and independent on the chaotic properties in the semiclassical limit. 
\par
The quite common phenomenon -  a constant and maximal EnPR equal to $h({\cal P})$ has been attributed in \cite{all} to free-independence of the dynamics $U$ and projections $P_j$.
The numerical results for our examples support this picture. Correlation functions of the order
$n$ are closer to zero (the value predicted by free-independence) for $n \leq 2\ln N/h({\cal P})$ than the values extrapolated from their long-time exponential decay. Moreover this exponential
decay can be easily explained by the hypothesis of the usual statistical independence for quantum variables  $F(m,j)=U^m P_j$, $m\neq 0$. Therefore, we observe that $\tau_{max}$ is a breaking time
from the free-independent behavior for $\{U, P_j\}$ to the independent one for $\{F(m,j)\}$, $m\neq 0$.
\par
\emph{Ackowledgements}  This work is supported by the grant 2P03B 084 25 of the Polish Ministry of Higher Education and Informatics.

\end{document}